\documentclass[russian,english]{article}
\usepackage[LGR,T1]{fontenc}
\usepackage[koi8-r,latin9]{inputenc}
\usepackage{booktabs}
\usepackage{textcomp}
\usepackage{graphicx}
\usepackage{setspace}
\usepackage{subscript}

\makeatletter

\DeclareRobustCommand{\greektext}{%
  \fontencoding{LGR}\selectfont\def\encodingdefault{LGR}}
\DeclareRobustCommand{\textgreek}[1]{\leavevmode{\greektext #1}}
\DeclareFontEncoding{LGR}{}{}
\DeclareTextSymbol{\~}{LGR}{126}
\providecommand{\tabularnewline}{\\}

\newcommand{\lyxaddress}[1]{
\par {\raggedright #1
\vspace{1.4em}
\noindent\par}
}

\makeatother

\usepackage{babel}
\begin{document}

\title{High-intensity pulsed ion beam treatment of amorphous iron-based
metal alloy\thanks{Preprint submitted to Journal of Physics: Conference Series (JPCS).
Work accepted at the Conference \textquotedbl{}Low temperature Plasma
in the Processes of Functional Coating Preparation\textquotedbl{},
Kazan, Russia, 2019.}}

\begin{singlespace}

\author{\noindent R.A.~Nazipov\textsuperscript{1,4}, R.I.~Batalov\textsuperscript{2},
R.M.~Bayazitov\textsuperscript{2},\inputencoding{koi8-r}\foreignlanguage{russian}{}\\
\inputencoding{latin9} H.A.~Novikov\textsuperscript{2}, V.A.~Shustov\textsuperscript{2},
E.N.~Dulov\textsuperscript{3}}
\end{singlespace}

\maketitle

\lyxaddress{\textsuperscript{1} Kazan National Research Technological University,
68 Karl Marx Str., 420015 Kazan, Russia }

\lyxaddress{\textsuperscript{2} Zavoisky Physical-Technical Institute, FRC Kazan
Scientific Center of RAS, 10/7 Sibirsky tract, 420029 Kazan, Russia }

\lyxaddress{\textsuperscript{3} Kazan Federal University, 18 Kremlyovskaya Str.,
420008 Kazan, Russia }

\lyxaddress{\textsuperscript{4} Author to whom any correspondence should be
addressed, rusnazipov@kstu.ru }
\begin{abstract}
The results of intense pulsed ion beam (IPIB) treatment of the soft
magnetic amorphous alloy of a FINEMET-type are presented. Foil produced
from the alloy was irradiated with short (about 100~ns) pulses of
carbon ions and protons with energy of up to 300~keV and an energy
density of up to 7~J/cm\textsuperscript{2}. X-ray diffraction, Mössbauer
spectroscopy and magnetic measurements were used to investigate structural
and magnetic properties of irradiated foils. It is shown that the
foil remains intact after the treatment, and the crystal structure
still amorphous. Spontaneous magnetization vector is found to lie
almost along perpendicular to the foil plane after irradiation, whereas
for the initial amorphous foil it belongs to the plane. The magnetic
properties of the foil undergo changes: the coercive force decreases,
the saturation induction increases slightly, and the magnetization
curve has shallower slope.
\end{abstract}

\section{Introduction }

Intense pulsed ion beams (IPIBs) with energy of 0.1-1~MeV generated
by a magnetically insulated ion diode are characterized to a greater
extent by the energy impact on the solid, since the ion fluence generated
in one pulse with duration of about 100~ns does not exceed 10\textsuperscript{13}~ion/cm2.
In this case, the ion current density can reach 250~A/cm\textsuperscript{2}
with the corresponding energy density up to 5-6~J/cm\textsuperscript{2}.
Such parameters of the ion beam provide surface melting and evaporation,
recrystallization, deformation, and destruction of materials \cite{key001,key002}. 

The effect of IPIBs on amorphous metal alloys (AMAs) has so far been
poorly studied and interesting from many points of view: the search
for promising radiation-resistant materials, the improvement of functional
properties, and fundamental researches on fast energy effects. IPIB
acts on the target for tens or hundreds of nanoseconds, and the main
energy of the ion beam is almost completely absorbed in a thin surface
layer (up to about 1~\textgreek{m}m) without causing strong heating
of the entire volume. 

AMAs are a relatively new class of materials that have a disordered
structure and unique properties related with this. They are obtained
by a combination of a special chemical composition and melt cooling
conditions. For iron-based alloys, metal-metalloid type chemical compositions
and ultrafast melt cooling at speeds of 10\textsuperscript{6}~K/s
are used. 

AMAs are metastable under normal conditions. When heated above a certain
temperature their properties change dramatically in a short time:
initially being solid, elastic and ductile, they become brittle. It
was previously shown \cite{key003} that AMA can be crystallized using
a flash light lamp with the radiation duration of about 0.1-1 ms.
The effect of a IPIB with a duration of about 100~ns on an amorphous
alloy has not been adequately studied \cite{key004,key005,key006,key007}.
Previous experiments with IPIBs irradiation with an energy density
of 1.5~J/cm\textsuperscript{2} showed \cite{key004,key005} that
the alloy did not crystallize, so it was necessary to irradiate AMA
with a maximum energy density.

\section{Experimental }

Iron-based AMA of FINEMET-type named 5BDSR (5NbCuSiB) with atomic
composition Fe\textsubscript{77}Cu\textsubscript{1}Nb\textsubscript{3}Si\textsubscript{13}B\textsubscript{6}
was used as samples. The amorphous alloy was produced as a ribbon
by Ashinsky plant using a single roller melt spinning technique. The
samples cut from the ribbon were pieces of foil with size 15\texttimes 10~mm
and thickness 25~micron. 

Samples were fixed in a steel holder with a mask with a hole of 10.5~mm
in diameter. During ion irradiation the energy density W varied from
1 up to 7~J/cm2 and the number of pulses N varied from 2 to 20. A
test irradiation of aluminum foil with a thickness of about 10~\textgreek{m}m
was also carried out. The parameters of the «TEMP» pulse ion accelerator
in which the irradiation was carried out were as follows: the composition
of the ion beam \textemdash{} Cn+ (80\%) and H+ (20\%); maximum ion
energy 300~keV; pulse duration of about 100~ns; the pressure in
the vacuum chamber is about 10\textsuperscript{-5}~mm~Hg. 

X-Ray diffraction (XRD) measurements of initial and irradiated samples
were carried out using Cu-K\textgreek{a} radiation both in standard
Bragg-Brentano geometry and in grazing incidence assymetric Bragg-Brentano
geometry. 

Mössbauer measurements were performed in transmission geometry with
a \textsuperscript{57}Co source in the rhodium matrix with an activity
of 50\textendash 40~mCi. Mössbauer measurements of the samples in
a magnetic field with strength of 2400~Oe were also made to exclude
the influence of the magnetic texture on the analysis of the magnetic
local structure. 

Magnetic properties were measured on an original setup based on LDC1000
EVM module, which measures the parameters of the oscillatory circuit
depending on the external magnetic field, in the inductance of which
the ferromagnetic sample under study is the core. The resonance frequency
of such a circuit is proportional to the differential magnetic susceptibility
of the sample, and a field dependence of the susceptibility is obtainable
as a result. The external magnetic field strength slowly changed in
the range of \textpm 1000~Oe during measurements.

\section{Results and Discussion }

After ion irradiation with the highest energy density W=5-6~J/cm\textsuperscript{2},
the samples remain elastic and are able to bend elastically under
load up to plastic deformation. They do not become brittle as after
conventional thermal annealing. Their shape changes, they bend in
the direction of the beam, probably under the influence of internal
stresses. Samples are not destroyed even with the number of pulses
N=20. The aluminum foil test sample is almost completely destroyed
in a single pulse with such an energy density. 

Samples after IPIBs irradiation remain X-ray amorphous both in volume
and on the surface. This is evidenced by the XRD patterns of the samples
with a grazing incidence angle \textgreek{f}=2\textdegree{} presented
in figure 1. XRD patterns obtained in standard Bragg-Brentano geometry
are not presented here because they are identical to the patterns
of the initial samples. 

\begin{figure}[h]
\begin{centering}
\includegraphics[scale=0.19]{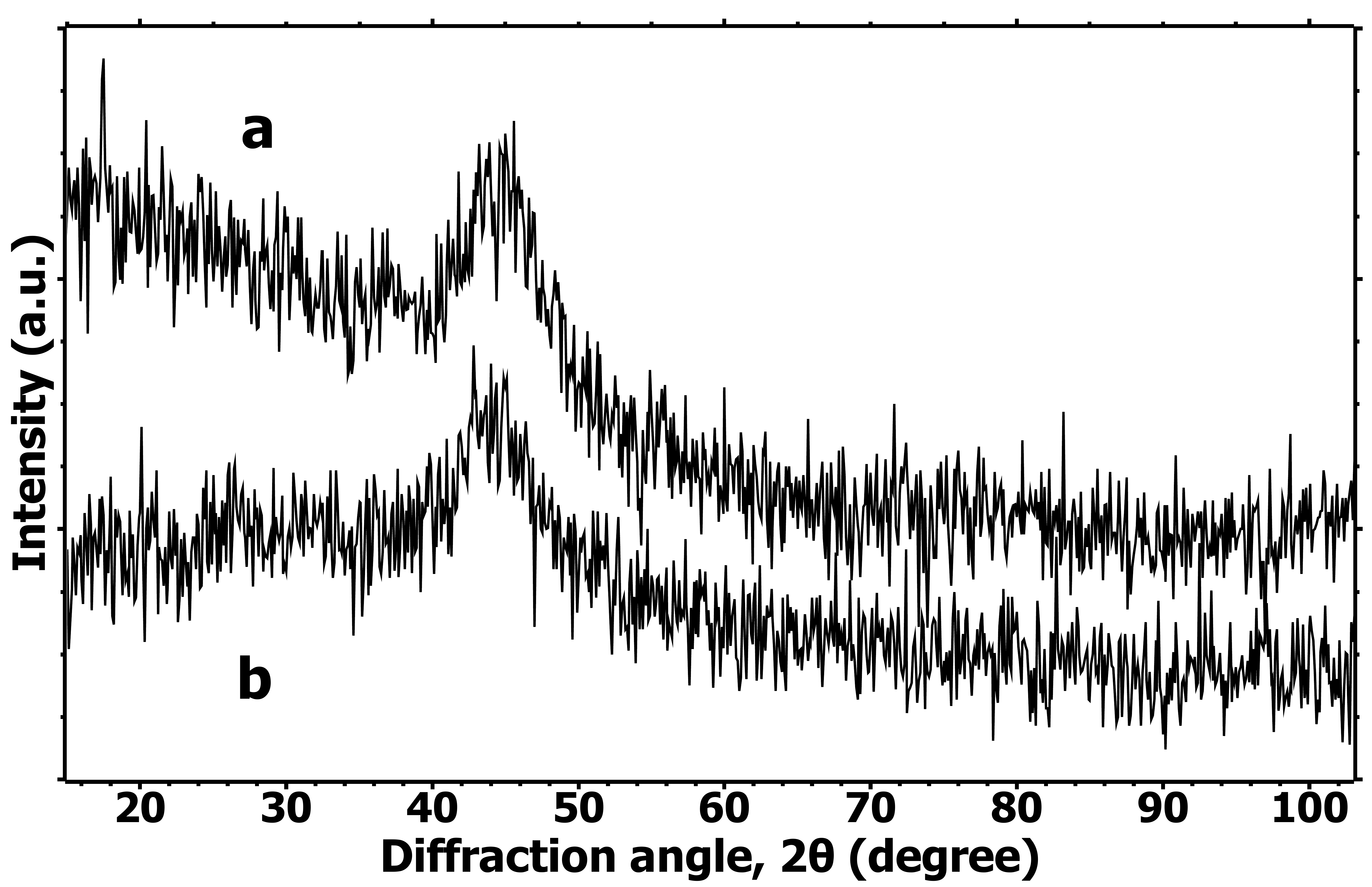}
\par\end{centering}

\caption{XRD patterns of the samples: a) the initial alloy; b) alloy after
irradiation (W=4.5~J/cm2, N=2).}

\end{figure}

Table 1 shows the parameters of the X-ray amorphous halo approximated
by a Gaussian line. For the samples irradiated with a high energy
density the X-ray amorphous halo shifts toward smaller angles 2\textgreek{j},
from 44.5\textdegree{} for the initial to 44.0\textdegree{} for the
irradiated. The full width at half maximum (FWHM) for the irradiated
sample becomes slightly greater.

\begin{table}[h]

\caption{X-ray amorphous halo parameters (irradiation with parameters W=4.5~J/cm\protect\textsuperscript{2},
N=2).}

\begin{centering}
\begin{tabular}{ccc}
\toprule 
Gaussian parameters & Initial sample & Irradiated sample\tabularnewline
\midrule
\midrule 
Center, 2\textgreek{j}\textdegree{} & 44.6 & 44\tabularnewline
\midrule 
Maximum, a.u. & 141.4 & 117.3\tabularnewline
\midrule 
Area, a.u. & 864  & 1061\tabularnewline
\midrule 
FWHM, 2\textgreek{j}\textdegree{} & 5.7  & 8.5\tabularnewline
\bottomrule
\end{tabular}
\par\end{centering}

\end{table}

Mössbauer measurements also confirm the amorphous nature of the samples.
The shape of the Mössbauer spectra is typical for disordered iron
compounds with an almost uniform distribution of nonequivalent positions
of Mössbauer nuclei in local atomic environment. There is a magnetically
splitted sextet consisting of wide spectral lines in all spectra. 

Figure 2 shows the Mössbauer spectra of the initial samples (a), the
irradiated samples (b) and the initial and irradiated samples measured
in a magnetic field (c). The difference between the irradiated and
non-irradiated samples is a change in the magnetic texture, since
the intensities of lines 2 and 5 of the wide sextet have changed.
The spontaneous magnetization vector becomes closer to the normal
to the foil plane in irradiated samples. This does not mean that the
sample was so magnetized, since it is magnetically soft. It means
that the induced magnetic anisotropy has become perpendicular. The
shape of the Mössbauer spectra does not depend on the number of pulses
of IPIB\textquoteright s irradiation.

\begin{figure}[h]
\begin{centering}
\includegraphics[scale=0.28]{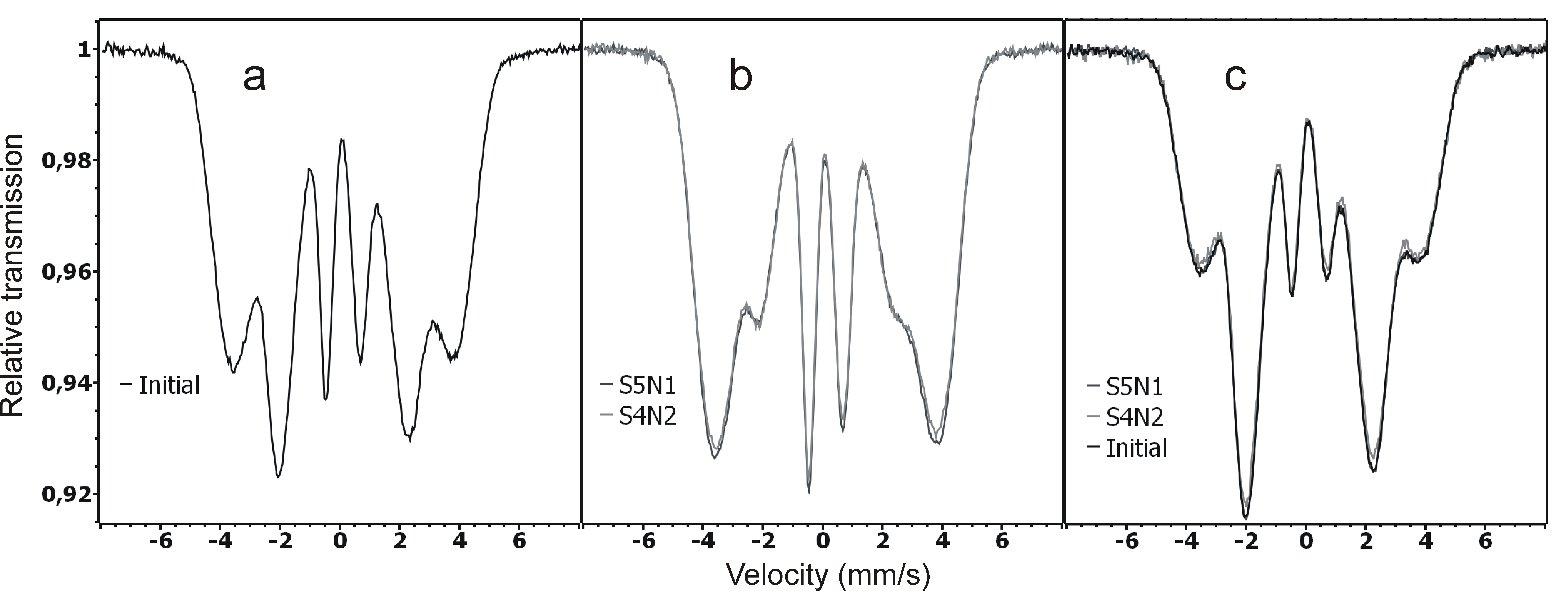}
\par\end{centering}

\caption{Mössbauer spectra of the amorphous alloy 5BDSR: a) initial sample;
b) samples after irradiation with W=5-6~J/cm\protect\textsuperscript{2},
S5N1 N=15, S4N2 N=4; c) the spectra of all samples measured in a magnetic
field.}

\end{figure}

The mathematical treatment of the Mössbauer spectra was carried out
with the restoration of the hyperfine field distribution function
P(H). Table 2 shows the found ratio of line intensities and the average
hyperfine field for the initial and irradiated samples.

\begin{table}[h]

\caption{Results of the mathematical treatment of Mössbauer spectra. The ratio
of intensities of second to first line of sextets denoted as $\frac{A2}{A1}$,
average hyperfine magnetic field denoted as $\overline{H}_{hf}$.
Pearson\textquoteright s chi squared test results are also presented
as $\chi^{2}$ (irradiation with parameters W=4.5~J/cm\protect\textsuperscript{2},
N=2; magnetic field 2.4~kOe).}

\centering{}%
\begin{tabular}{ccccccc}
\toprule 
 & \multicolumn{3}{c}{No external magnetic field} & \multicolumn{3}{c}{In external magnetic field}\tabularnewline
\midrule 
\addlinespace
 & $\chi^{2}$ & $\frac{A2}{A1}$ & $\overline{H}_{hf}$, kOe & $\chi^{2}$ & $\frac{A2}{A1}$ & $\overline{H}_{hf}$, kOe\tabularnewline\addlinespace
\midrule 
Initial sample & 3.94 & 0.71 & 215 & 2.59 & 1.25 & 212\tabularnewline
\midrule 
Irradiated sample & 5.68 & 0.29 & 221 & 2.74 & 1.25 & 212\tabularnewline
\bottomrule
\end{tabular}
\end{table}

It can be seen from this table that the average hyperfine field of
the irradiated sample is slightly greater than for the initial one,
which could be explained by a decrease in the free volume \cite{key008}.
However, measurements of the Mössbauer spectra in a magnetic field
showed that the effect of increasing the field is the influence of
the magnetic texture. The spectra with the same direction of the magnetic
moment of the nucleus are almost indistinguishable (figure 2c). 

Despite weak changes in the structure of the alloy shown by XRD and
Mössbauer studies, the magnetic properties and the eddy current magnetic
impedance (figure 3a) vary significantly for the irradiated samples
in comparison with the initial ones. 

By integrating the field dependences of the differential magnetic
susceptibility, we can obtain the dependences of the magnetizations
on the magnetic field (hysteresis loops, figure 3b). A comparison
of the hysteresis loops of the initial and irradiated samples showed
that the magnetic anisotropy in the samples changes, the coercive
field decreases, and the saturation magnetization increases as a result
of irradiation. In the longitudinal orientation of the sample, when
the lines of force of the magnetic field coincide with the plane of
the sample and with the direction of the ribbon casting, the initial
sample reaches saturation at lower fields (about 200~Oe) than the
irradiated one (about~400 Oe).

\begin{figure}[h]
\begin{centering}
\includegraphics[scale=0.28]{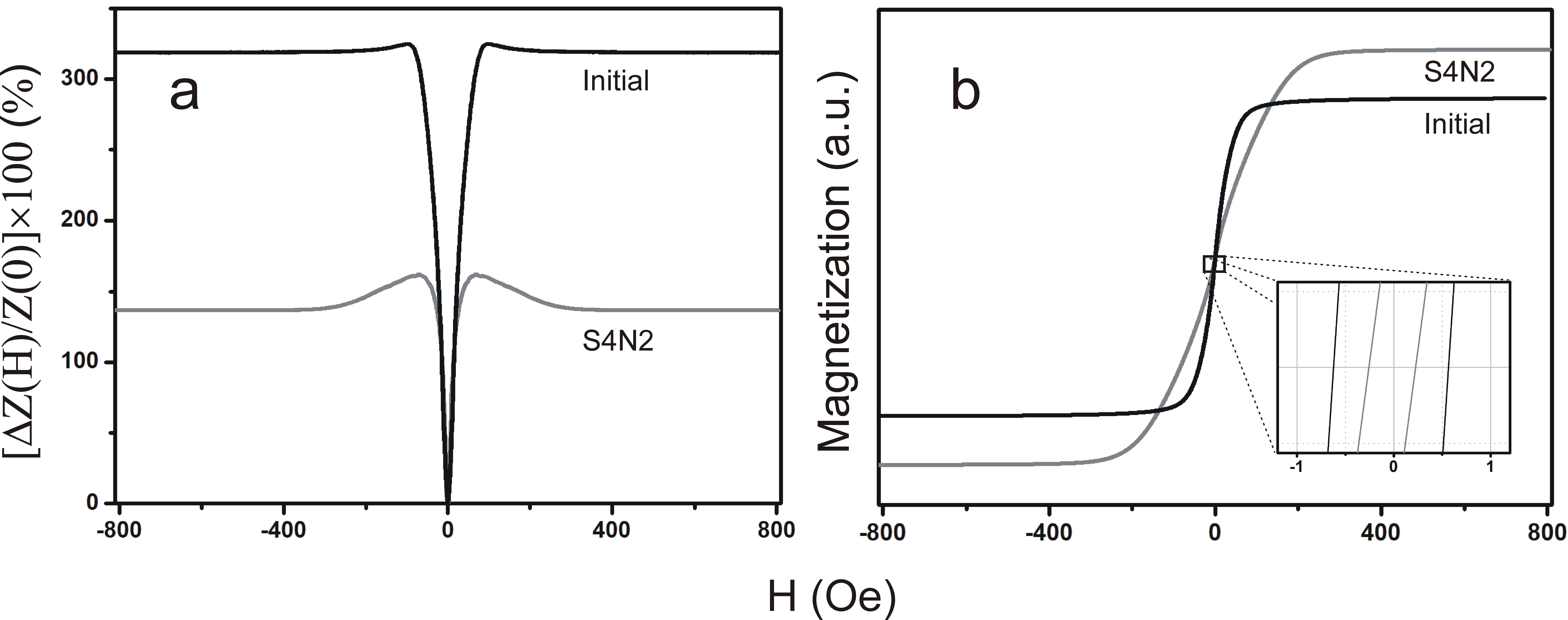}
\par\end{centering}

\caption{Figure 3. Magnetic properties of the samples: a) relative eddy current
magnetic impedance depending on the external magnetic field H, here
\textgreek{D}Z=Z(H)-Z(0) is the difference between zero-field and
in-field impedances; b) hysteresis loops obtained by integrating the
ACMS data. The inset on the right graph shows enlarged low-field region
of the magnetization curve.}

\end{figure}

The field dependence of the parameters of the oscillatory circuit
was measured at frequencies of 1-2~MHz. Used technique of the alternating
current magnetic susceptibility measurement (ACMS) caused certain
increase of the observed coercive force. Therefore, the values obtained
in present work should differ from the well-known ones measured in
a constant magnetic field. However, this data can be used to compare
values of the coercive field of different samples. Since the samples
are metallic and have good electrical conductivity, the alternating
field of the inductor creates eddy currents in them. These eddy currents
in turn affect the complex resistance of the measuring oscillatory
circuit. Eddy currents penetrate to the skin depth, which depends
on the values of magnetic susceptibility. Thus, the value of the complex
resistance of the oscillatory circuit, which contains the sample under
study as a core of the inductance, depends on the strength of the
external scanning magnetic field. The dependence of the complex resistance
of the measuring circuit on the magnetic field represents the eddy
current magnetic impedance. Typically, the magnetic impedance of amorphous
alloys in the form of ribbons or wires is observed using a galvanic
connection. In our case, the magnetic impedance is observed without
galvanic contact, but by means of the eddy currents. 

In the irradiated samples the relative values of the eddy current
impedance decreases more than two times compared with the initial
ones (figure 3a).

\section{Conclusion }

On the basis of the presented results, the amorphous FINEMET-type
alloy remains amorphous after irradiation with intense pulsed ion
beams (IPIBs). Internal mechanical stresses occur in the alloy after
irradiation. 

The value of the average hyperfine magnetic field for samples with
applied external magnetic field 2400~Oe does not change within the
measurement error. Magnetic properties are changed for modified samples.
Magnetic anisotropy changes, the coercive field decreases (2\textendash 2.5~times),
and the saturation magnetization slightly increases.

\section*{Acknowledgements }

The device for measuring magnetic properties was developed and created
at the support of the subsidy allocated to Kazan Federal University
for the state assignment in the sphere of scientific activities (3.7352.2017/8.9).\\

\noindent This work was supported by the Russian Foundation for Basic
Research (Grant No. 19-03-00847).

\end{document}